\documentclass[english,aps,manuscript,showpacs,superscriptaddress]{revtex4}
\usepackage[T1]{fontenc}
\usepackage[latin9]{inputenc}
\usepackage{verbatim}
\usepackage{amsmath}
\usepackage{amssymb}
\usepackage{graphicx}
\usepackage{esint}
\usepackage{textcomp}
\usepackage{color}

\makeatletter
\@ifundefined{textcolor}{}
{%
 \definecolor{BLACK}{gray}{0}
 \definecolor{WHITE}{gray}{1}
 \definecolor{RED}{rgb}{1,0,0}
 \definecolor{GREEN}{rgb}{0,1,0}
 \definecolor{BLUE}{rgb}{0,0,1}
 \definecolor{CYAN}{cmyk}{1,0,0,0}
 \definecolor{MAGENTA}{cmyk}{0,1,0,0}
 \definecolor{YELLOW}{cmyk}{0,0,1,0}
}

\newif\ifhyper
\hypertrue
\ifhyper       

\usepackage{babel}

\makeatother

\usepackage{babel}
\begin{document}

\title{The Local Potential Approach to frustrated antiferromagnets}

\author{Shunsuke Yabunaka}

\affiliation{Yukawa Institute for Theoretical Physics, The Kyoto University, Kitashirakawa
Oiwake-Cho, 606-8502 Kyoto, Japan}

\author{Bertrand Delamotte}

\affiliation{Sorbonne Universit\'{e}s, UPMC Univ Paris 06, LPTMC, CNRS UMR 7600, F-75005, Paris, France}

\begin{abstract}
We revisit the critical behavior of classical frustrated systems using the nonperturbative renormalization group (NPRG) 
equation. Our study is performed within the local potential approximation of this equation
to which is added the flow of the field renormalization. Our flow equations are functional
to avoid possible artifacts coming from field expansions which consists in keeping only  
a limited number of coupling constants.  We present a simple numerical method to follow the fixed point solution of our equations
by changing 
gradually the dimension $d$ and the  number $N$ of spin-components. We explain in details the advantage of this method as well as
the numerical difficulties we encounter, which become severe close to 
$d=2$. The function $N_c(d)$ separating the regions of first and second order in the $(d,N)$ plane is computed for $d$ between 4 and 2.2. 
Our results confirm what was previously found within cruder approximation of the  NPRG equation
and contradict both the fixed dimension perturbative approach and the results
obtained within the conformal bootstrap approach.
\end{abstract}
\maketitle

\section{Introduction}
The critical behavior of antiferromagnetic frustrated systems is still a debated question  
forty years after the first studies of these systems \cite{Delamotte Review,Kawamura Review}. The key difference between 
frustrated and nonfrustrated systems is that the order parameter
is a vector in the nonfrustrated case and a matrix in the other cases. When frustration 
originates from the geometry of the system as in Stacked triangular Antiferromagnets (STA),
the symmetry of the Hamiltonian is $O(N)\otimes O(2)$ for $N$-component spins and the order
parameter is a rectangular $N\times 2$ matrix \cite{Order parameter symmetry}. Depending on $N$ and the dimension $d$ of space,
the nature of the phase transition changes, being first order for low values of $N$ and dimensions
close to four and second order otherwise. One of the key questions is thus the determination of the 
line $N_c(d)$ separating the first and second order  regions. It turns out that the value of $N_c(d=3)$
is certainly close to 3 and its precise determination is crucial to know whether the transition is
first or second order for the systems realized in nature that are either Ising, XY or Heisenberg. 
Numerical simulations of several frustrated antiferromagnets such as XY and Heisenberg STA 
show unambiguously that the transition is first order for these systems \cite{Loison1, Loison2, STA Itakura, Tanh Ngo Diep}.
However, depending on the theoretical approach considered, the determination 
of $N_c(d)$ varies much when $d\lesssim 3.3$
 and, as a result, it is not yet settled whether all $O(N)\otimes O(2)$ symmetric systems 
 undergo first order phase transitions in $d=3$ for $N\le3$. The two-dimensional physics
 of the XY and Heisenberg systems is also debated because the relevance of topological defects
 is not yet understood, in particular the possibility that they trigger a phase transition at finite
 temperature \cite{Kawamura Miyashita, Wintel, Stephan, Caffarel, Calabrese focus fixed points 2, Azaria}.
 
The different theoretical approaches tackling with the problem of the calculation of $N_c(d)$
can be roughly divided into two classes:  the perturbative and the nonperturbative renormalization group (NPRG)
calculations.
The class of perturbative calculations can be again divided into several different subclasses
depending on whether they are performed directly in $d=3$ (at six loops) \cite{Pelissetto,Calabrese focus fixed points,Calabrese 3} or in an $\epsilon$-
or pseudo-$\epsilon$-expansion (respectively at six and five loops)\cite{Calabrese five-loop}. In the latter case, the value of $N_c(d=3)$
is systematically found larger than 3 (of order 6) as it is also the case for the NPRG calculations
that find $N_c(d=3)\simeq 5.1$\cite{Zumbach1, Zumbach2, NPRG field expansion, Delamotte Review, NPRG semi-expansion}. On the contrary, the perturbative calculation performed directly in
$d=3$ at six loops yields a fixed point for $N=2$ and 3 and thus predicts that several $O(N)\otimes O(2)$ symmetric
systems should undergo a second order phase transition. 

Recently, a completely different method based on the conformal bootstrap has been used to study matrix
models in $d=3$ and in particular the $O(N)\otimes O(2)$ frustrated systems \cite{Nakayama Ohtsuki-1, Nakayama Ohtsuki-2}. A critical behavior has been
found in the Heisenberg case with exponents in good agreement with those of the six-loop fixed
dimension approach. This approach has the advantage of being unbiased by convergence 
problems since it is not based on series expansions, contrary to RG methods and, when applied to
the ferromagnetic $O(N)$ models, it leads to an extremely accurate determination of the critical
exponents, at least when it is truncated at large orders \cite{Bootstrap Ising, Bootstrap Ising-2, Bootstrap O(N), Bootstrap O(N) 2}.

The situation of the NPRG approach, that we re-examine here, is therefore the following. Either 
the conclusions drawn from its results are correct and then both the fixed dimension perturbative
RG approach and the conformal bootstrap are wrong or, conversely, it is wrong (together with the 
$\epsilon$-expansion approaches) and this implies that
the approximations used are too drastic to reproduce the correct physics. In both cases, something very unusual
is at work because the methodologies that have been used in these studies 
lead in many cases to correct and accurate results.

As for the NPRG, which is based on an exact RG equation, the approximations used so far
to tackle with frustrated systems consists in performing a derivative expansion \cite{Berges} and a field expansion
of the Gibbs free energy \cite{NPRG field expansion, Delamotte Review, NPRG semi-expansion}. The rationale behind this choice is  (i) that the critical behavior of thermodynamic quantities
such as the specific heat or the susceptibility for instance are dominated by long wavelength fluctuations
which justifies expanding the correlation functions in their momenta (derivative expansion) and (ii) 
that the impact of the $n$-point functions with $n$ large on the RG flow
of the zero or two-point functions should be small (field-expansion). It is the aim of this article to
eliminate one source of inaccuracy of the NPRG approach, the field expansion, which is known to be inaccurate
at low dimensions even for simple models such as the ferromagnetic $O(N)$ models \cite{Delamotte Review}. The price to pay to get
rid of this approximation is to work functionally, that is, to follow the RG flow of functions of
the fields instead of a limited number of coupling constants. In the case of nonfrustrated systems,
this is relatively simple since the $O(N)$ symmetry implies that all functions involved in 
the RG flows depend on the fields
only through the unique $O(N)$-invariant: $\rho={\vec\phi}^{\,2}$. For frustrated systems,
there exists two $O(N)\otimes O(2)$ invariants and the resulting flow equations are partial differential
equations that are rather involved. We show in this article how to simplify the numerical problem
and point out why the  numerical difficulties are so severe at low dimensions that our method does no 
longer work when approaching $d=2$. We provide the results thus obtained for the curve $N_c(d)$ 
between $d=4$ and $d=2.2$. Our results confirm what was previously found within a NPRG approximation involving a  field expansion of the potential
and the $\epsilon$-approaches and thus contradict both the fixed-dimension perturbative approach and the results
obtained with the conformal bootstrap.

\section{The Model}

As the archetype of frustrated spin systems, we employ the Stacked
Triangular Antiferromagnets (STA). This system is composed of two-dimensional
triangular lattices that are piled-up in the third direction. At each
lattice site $i$, is defined a  $N$-component
vector $\mathbf{S}_{i}$ of modulus 1.  The Hamiltonian of this system
is given by

\begin{equation}
H=\sum_{\left\langle ij\right\rangle }J_{ij}\mathbf{S}_{i}\cdot\mathbf{S}_{j}.
\end{equation}
The sum $\left\langle ij\right\rangle $ runs on all pairs of nearest
neighbor spins. The coupling constants $J_{ij}$ are given by $J_{\perp}$
for a pair of sites inside a plane and $J_{\parallel}$ between planes.
We assume that the interactions inside a plane are antiferromagnetic:
$J_{\perp}$ is positive.

The long distance effective theory for the STA has been derived by
Yosefin and Domany\cite{Order parameter symmetry}. The order parameter consists of
the $N\times2$ matrix $\Phi=\left(\mathbf{\boldsymbol{\phi}}_{1},\mathbf{\boldsymbol{\phi}}_{2}\right)$
that satisfies 
\begin{equation}
\boldsymbol{\phi}_{i}\cdot\boldsymbol{\phi}_{j}=\delta_{ij}
\end{equation}
for $i,j=1,2$. Then, the effective Hamiltonian in the continuum is
given by 
\begin{equation}
H=\int d^{d}\mathbf{x}\left(\frac{1}{2}\left[\left(\partial\boldsymbol{\phi}_{1}\right)^{2}+
\left(\partial\boldsymbol{\phi}_{2}\right)^{2}\right]\right).
\end{equation}

The constraint $\boldsymbol{\phi}_{i}\cdot\boldsymbol{\phi}_{j}=\delta_{ij}$
for $i,j=1,2$ can be replaced by a soft potential 
$U\left(\mathbf{\boldsymbol{\phi}}_{1},\mathbf{\boldsymbol{\phi}}_{2}\right)$
whose minima are given by $\boldsymbol{\phi}_{i}\cdot\boldsymbol{\phi}_{j}=const\times\delta_{ij}$
and the Ginzburg-Landau-Wilson Hamiltonian for STA reads 
\begin{equation}
H=\int d^{d}\mathbf{x}\left(\frac{1}{2}\left[\left(\partial\boldsymbol{\phi}_{1}\right)^{2}+
\left(\partial\boldsymbol{\phi}_{2}\right)^{2}\right]+U\left(\mathbf{\boldsymbol{\phi}}_{1},\mathbf{\boldsymbol{\phi}}_{2}\right)\right).\label{eq:effective hamiltonian}
\end{equation}
Instead of $\mathbf{\boldsymbol{\phi}}_{i}$, it is convenient to
work with the invariants of the O$(N)\times$O(2) group that can be
chosen as: 
\begin{equation}
\begin{array}{ll}
\rho= & \mathrm{Tr}\left(^{t}\Phi\Phi\right)=\boldsymbol{\phi}_{1}^2+ \boldsymbol{\phi}_{2}^2,\\
\tau= & \frac{1}{2}\mathrm{Tr}\left(^{t}\Phi\Phi-\rho/2\right)^{2}=
\frac{1}{4}\left(\boldsymbol{\phi}_{1}^2- \boldsymbol{\phi}_{2}^2\right)^2
+\left(\boldsymbol{\phi}_{1}. \boldsymbol{\phi}_{2}\right)^2.
\end{array}
\end{equation}
With this choice, the ground state configuration corresponds to $\rho={\rm const.}$
and $\tau=0$.  Up to the fourth order $U\left(\rho,\tau\right)$
can be written as  
\begin{equation}
U\left(\rho,\tau\right)=\frac{\lambda}{2}\left(\rho-\kappa\right)^{2}+\mu\tau,
\label{pot}
\end{equation}
where $\lambda$ and $\mu$ are positive coupling constants. A typical
ground state in terms of $\Phi$  is
given by $\Phi_{\alpha,i}=\sqrt{\kappa/2}\delta_{\alpha,i}$, that is:
\begin{equation}
\Phi_{\rm min}\equiv\left(\begin{array}{cc}
\sqrt{\frac\kappa 2} & 0\\
0 & \sqrt{\frac\kappa 2}\\
\vdots & \vdots\\
0 & 0
\end{array}\right).
\label{eq:min}
\end{equation}

\section{The nonperturbative renormalization group equation}

The NPRG method is based on Wilson's idea of integrating statistical
fluctuations step by step. In this paper, we employ the effective
average action method as an implementation of the NPRG in continuum
space \cite{Wetterich1,Ellwanger,Morris, Wetterich2}.

 The first step is to
introduce a $k$-dependent partition function $\mathcal{Z}_{k}$ in
the presence of sources: 
\begin{equation}
\mathcal{Z}_{k}\left[\boldsymbol{J}_{i}\right]=
\int\mathcal{D}\boldsymbol{\phi}_{i}\exp\left(-H[\boldsymbol{\phi}_{i}]-
\Delta H_{k}[\boldsymbol{\phi}_{i}]+\boldsymbol{J}_{i}\cdot\boldsymbol{\phi}_{i}\right),
\end{equation}
{where $\mathbf{J}_{i}\cdot\boldsymbol{\phi}_{i}=
\sum_{i=1}^{2}\int_{x}\mathbf{J}_{i}\left(\mathbf{x}\right)\cdot\boldsymbol{\phi}_{i}\left(\mathbf{x}\right),$
and $\Delta H_{k}=\sum_{i=1}^{2}\boldsymbol{\phi}_{i}(x) R_{k}(x-y)\boldsymbol{\phi}_{i}(y)$.}
 The idea underlying the effective average action is that
in $\mathcal{Z}_{k}$ only the fluctuations of large wave-numbers
(the rapid modes) compared to $k$ are integrated over while  the others (the slow modes)
are frozen by the $\Delta H_{k}$ term. As $k$ is decreased, more and more modes are integrated
until they are all when $k=0$. The function $R_{k}({q}^{2})$, which is the
Fourier transform of $R_{k}({x})$, plays the role of separating rapid
and slow modes: It almost vanishes for $\vert q\vert>k$ so that the
rapid modes are summed over and is large (of order $k^{2}$) below
$k$ so that the fluctuations of the slow modes are frozen. We define
as usual $W_{k}[\boldsymbol{J}_{i}]=\ln\mathcal{Z}_{k}[\boldsymbol{J}_{i}]$.
 Thus, the order parameter $\boldsymbol{\varphi}_{j}\left(\mathbf{x}\right)$
at scale $k$ is defined by 
\begin{equation}
{\boldsymbol{\varphi}_{i}\left(\mathbf{x}\right)}=\left\langle \boldsymbol{\phi}_{i}
\left(\mathbf{x}\right)\right\rangle =
\frac{\delta W_{k}\left[\boldsymbol{J}_{i}\right]}{\delta\boldsymbol{J}_{i}\left(\mathbf{x}\right)}\, .
\label{eq:def=00003D00003D0000A5phi}
\end{equation}
The running effective average action $\Gamma_{k}\left[\mathbf{\boldsymbol{\varphi}}_{i}\right]$
is defined  as the (modified) Legendre transform of $W_{k}$:
\begin{equation}
\Gamma_{k}\left[\mathbf{\boldsymbol{\varphi}}_{i}\right]=-W_{k}\left[\boldsymbol{J}_{i}\right]+\mathbf{J}_{i}\cdot\boldsymbol{\varphi}_{i}-\Delta H_{k}\left[\mathbf{\boldsymbol{\varphi}}_{i}\right]
\end{equation}
where $\boldsymbol{J}_{i}$ is defined such that Eq. (\ref{eq:def=00003D00003D0000A5phi})
holds for fixed $\mathbf{\boldsymbol{\varphi}}_{i}$. From this definition
one can show that 
\begin{equation}
\begin{cases}
\Gamma_{k=\Lambda}\simeq H\\
\Gamma_{k=0}=\Gamma
\end{cases},\label{gammakboundaries}
\end{equation}
where the cutoff $\Lambda$ is the inverse of the lattice spacing
$a$. Equations (\ref{gammakboundaries}) imply that $\Gamma_{k}$
interpolates between the Hamiltonian of the system when no fluctuation
has been summed over, that is, when $k=\Lambda$, and the Gibbs free
energy $\Gamma$ when they have all been integrated, that is, when
$k=0$. We define the variable $t$, called ``RG time'', by $t=\ln\left(k/\Lambda\right)$.
The exact flow equation for $\Gamma_{k}$ reads \cite{Wetterich1,Wetterich2}:
\begin{equation}
\partial_{t}\Gamma_{k}[\mathbf{\boldsymbol{\varphi}}_{i}]=
\frac{1}{2}\mathrm{Tr}\int_{x,y}\partial_{t}R_{k}(x-y)\left(\frac{\delta^{2}\Gamma_{k}
\left[\mathbf{\boldsymbol{\varphi}}_{i}\right]}{\delta\varphi_{i}^{\alpha}
\left(\mathbf{x}\right)\delta\varphi_{i'}^{\alpha'}\left(\mathbf{y}\right)}+R_{k}
\left(\mathbf{x-y}\right)\delta_{i,i'}\delta_{\alpha,\alpha'}\right)^{-1},
\label{floweq}
\end{equation}
for $\alpha,\alpha'=1,2,\cdots N$ and $i,i'=1,2$.

\section{Truncations of the NPRG equation}

It is generally not possible to solve exactly the above flow equation (\ref{floweq})
and  approximations are required in practice. In this paper,
we employ the  approximation of lowest level in the derivative expansion
dubbed the local potential approximation (LPA) and some of its refinements.

Within the LPA, $\Gamma_{k}$ is approximated by a series expansion
in the gradient of the field, truncated at its lowest non trivial
order: 
\begin{equation}
\Gamma_{k}\left[\mathbf{\boldsymbol{\varphi}}_{i}\right]=
\int d^{d}\mathbf{x}\left(\frac{1}{2}\left[\left(\partial\boldsymbol{\varphi}_{1}\right)^{2}+
\left(\partial\boldsymbol{\varphi}_{2}\right)^{2}\right]+U_{k}\left(\rho,\tau\right)\right).
\end{equation}
Only a potential term $U_{k}\left(\rho,\tau\right)$ is thus retained in this approximation
which is accurate as long as the impact of the renormalization of the derivative terms on the flow of the potential
is small. This is most probably the case when the anomalous dimension is small and $d>2$. The next level
of approximation consists in including in the approximation a running field renormalization 
$Z_k$ 
\begin{eqnarray}
\Gamma_k=\int_x \Big\{U_k(\rho,\tau)+\frac{1}{2}
Z_k\Big(\big(\partial \vec \varphi_1\big)^2+ \big(\partial \vec
\varphi_2\big)^2\Big)
\label{action_generale2}
\end{eqnarray}
This approximation  has been used in \cite{tissier00b,tissier01, tissier03,  NPRG field expansion, Delamotte Review, NPRG semi-expansion}  
where  the function $U_k(\rho,\tau)$ was further expanded in powers of the invariants $\rho$ and $\tau$. 
This is what we  improve here to avoid any artifact coming from this field truncation.
This approximation, that we call LPA', yields the one-loop result obtained  
within the $\epsilon$-expansion in   $d=4-\epsilon$ and also, in the $O(N)$ case,
the one-loop result of the $\epsilon=d-2$ expansion of the nonlinear sigma model.
Although the situation is a little more involved in our case, it is very probable that 
the LPA' is very accurate close to $d=2$ and our numerical results confirm this, see the following.
Our approach is therefore at least a clever interpolation
between the results obtained either in $d=4$ or $d=2$.

The $k$-dependent effective potential $U_{k}\left(\rho,\tau\right)$
is defined by
\begin{equation}
\Omega U_{k}\left(\rho,\tau\right)=\Gamma_{k}\left[\mathbf{\boldsymbol{\varphi}}_{i}\right]
\end{equation}
where $\boldsymbol{\varphi}_{i},i=1,2$ are constant fields and $\Omega$
is the volume of the system. The running field renormalization $Z_{k}$ 
is set to one in LPA: $Z_{k}^{{\rm LPA}}=1$, which leads
to a vanishing anomalous dimension: $\eta=0$. In LPA' calculations,
the anomalous dimension $\eta$ is obtained from the flow of $Z_{k}$
since it can be shown that at criticality: 
\begin{equation}
Z_{k\rightarrow0}\sim\left(\frac{k}{\Lambda}\right)^{-\eta}.
\end{equation}
The flows of $U_k$ and $Z_k$ 
have been derived in \cite{NPRG field expansion, Delamotte Review, NPRG semi-expansion} and we recall them for completeness in 
Appendix A.
These flows are rather complicated and their numerical integration suffers from all the inherent
difficulties of the nonlinear partial differential equations. 

The first difficulty comes from the choice of variables. 
It is tempting to work with the
invariants $\rho$ and $\tau$ defined above because the symmetry of the problem is encoded 
in the very definition of the variables and any smooth function of these variables corresponds to 
a function that has the right symmetry. However, $\rho$ and $\tau$ satisfy $\frac{1}{4}\rho^{2}\geq\tau\geq0$
and it is not easy to deal with this constraint numerically because the domain where the
variables $\rho$ and $\tau$ live is nontrivial. Thus, we define another
set of variables $\psi_{i}$ which is numerically more convenient.
For any $\boldsymbol{\varphi}_{1}$ and $\boldsymbol{\varphi}_{2}$,
it can be proven that there exists $O_{1}\in O\left(N\right)$ and
$O_{2}\in O\left(2\right)$ such that the matrix $M\equiv O_{1}\Psi O_{2}$
,{ where $N\times2$ matrix $\Psi$ is defined as
$\Psi=\left(\mathbf{\boldsymbol{\varphi}}_{1},\mathbf{\boldsymbol{\varphi}}_{2}\right)$,}
becomes ``diagonal'', namely, 
\begin{equation}
M\equiv\left(\begin{array}{cc}
\psi_{1} & 0\\
0 & \psi_{2}\\
\vdots & \vdots\\
0 & 0
\end{array}\right).
\end{equation}
Because of the  $O\left(N\right)\times O\left(2\right)$ symmetry of the model, we conclude
that $U_{k}\left(\Psi\right)=U_{k}\left(M\right)$. This fact shows
that we can parametrize the order parameter space using $\psi_{1}$
and $\psi_{2}$, instead of $\boldsymbol{\varphi}_{1}$ and $\boldsymbol{\varphi}_{2}$.
The $O\left(N\right)\times O\left(2\right)$ invariants $\rho$ and
$\tau$ are expressed in terms of $\psi_{1}$ and $\psi_{2}$ as 
\begin{equation}
\begin{array}{ll}
\rho & =\psi_{1}^{2}+\psi_{2}^{2}\\
\tau & =\frac{1}{4}\left(\psi_{1}^{2}-\psi_{2}^{2}\right)^{2}.
\end{array}\label{eq:def_rho_tau}
\end{equation}
 From the definitions (\ref{eq:def_rho_tau}) we find that the symmetries of the original problem imply:  
\begin{equation}
U_{k}\left(\psi_{1},\psi_{2}\right)=U_{k}\left(-\psi_{1},\psi_{2}\right)=
U_{k}\left(\psi_{1},-\psi_{2}\right)=U_{k}\left(\psi_{2},\psi_{1}\right).\label{constraint}
\end{equation}
 Thus, to solve the flow equations, it is sufficient to consider the region $\psi_{2}\geq\psi_{1}\geq0$.
This triangular domain is much more convenient from a numerical
point of view than the parabolic domain $\frac{1}{4}\rho^{2}\geq\tau\geq0$
for the invariants $\rho$ and $\tau$.

At criticality, the $k$-dependent effective action is attracted towards
the fixed point solution of the NPRG equation once it is expressed
in terms of the dimensionless renormalized fields $\tilde{\psi}_{i}$
and a dimensionless local potential $\tilde{U}_{k}(\tilde{\psi}_{i})$.
We thus define the dimensionless and renormalized quantities: 
\begin{equation}
\begin{array}{l}
\tilde{\psi}_{i}=\left(Z_{k}k^{2-d}\right)^{1/2}\psi_{i}\\
\tilde{U}_{k}(\tilde{\psi}_{i})=k^{-d}U_{k}\left(\psi_{i}\right).
\end{array}
\end{equation}
The flow equation for $\tilde{U}_{k}$ is given by Eq. (\ref{eq:flowU})
in Appendix A. The critical exponent $\nu$ of the correlation length
is obtained from the relevant eigenvalue of the linearized flow around
the fixed point solution and $\eta$ from the flow of $Z_{k}$. The
other critical exponents can be deduced from these ones by scaling
relations.

The scaled $O\left(N\right)\times O\left(2\right)$ invariants $\tilde{\rho}$
and $\tilde{\tau}$ are defined by $\tilde\rho=Z_k k^{2-d}\rho$, $\tilde\tau=Z^2_k
k^{2(2-d)}\tau$, and the potential and couplings by $\widetilde{U}_k(\tilde\rho,\tilde\tau)=k^{-d}{U}_k(\rho,\tau)$,  
$y=q^2/k^2$,  $R_k(q^2)=Z_k k^2 y r(y)$. Notice that as said above
$Z_k$ does not reach a fixed point but  $\eta_k$, defined by $\eta_k=-d\log Z_k/d\log k$, does: 
$\eta_{k\to0}\to \eta$ at criticality with $\eta$ the anomalous dimension of the fields. 

\section{Numerical methods}

\subsection{The fixed point}

From a numerical point of view, there are two possibilities for finding
fixed points when they exist. The first is to dynamically integrate the flow. In this
case, the problem is to find the critical surface which is usually
done by dichotomy on the temperature. Once it is found, the fixed point is (approximately)
reached since it is attractive on the critical surface. The other
method is to look directly for the solution of the fixed point equation
(coupled with Eq. (\ref{eq:eta})): 
$\partial_{t}\tilde{U}^{*}(\tilde{\psi}_{i})=0$.
This is what we do here. The advantage of this method is three-fold:
(i) The numerical scheme is much simpler than integrating the flow; 
(ii) several numerical instabilities
occuring during the integration of the flow are avoided; (iii) the
critical exponents are easily obtained from the diagonalization of
the RG flow around the fixed point. We show in the following that 
although this scheme works very well in dimension $d=3$, numerical 
difficulties appear in dimensions close to $d=2$ that make almost impossible to
study the physics of frustrated systems in this dimension, at least with our numerical scheme.

The basic idea of this scheme is simple. It consists in solving the fixed point equations for $\tilde{U}^{*}$
on a grid in $(\psi_{1},\psi_{2})$ space, taking into account
the symmetries (\ref{constraint}) of this space. We introduce a cut-off field value $\tilde{\psi}_{max}$
and consider the domain $D:\tilde{\psi}_{max}\geq\tilde{\psi}_{2}\geq\tilde{\psi}_{1}\geq0$.
We then discretize $D$ on a square lattice with mesh size $\Delta\tilde{\psi}=\tilde{\psi}_{max}/\left(N_{p}-1\right)$,
where $N_{p}$ is the number of lattice points on the axis $\psi_{1}=0$.
The lattice points are given by $\left(i\Delta\tilde{\psi},j\Delta\tilde{\psi}\right)$
for integers $i$ and $j$ that satisfy $0\leq i\leq j\leq N_{p}-1$.
We define $\tilde{U}_{t}\left(i,j\right)\equiv\tilde{U}_{t}\left(i\Delta\psi,j\Delta\psi\right)$
to alleviate the notation.

The fixed point equation for the potential is a differential equation. We transform
it into a set of algebraic equations by discretizing the derivatives of $\tilde{U}$.
We give below some details about this procedure because all our numerical
problems come from the boundary of the domain $D$, precisely at the
points where the discretization involves exceptional cases.

The formulae for the derivatives $\tilde{U}_{t}^{\left(l,m\right)}\left(i,j\right)$
for $l,m=0,1,2$ are constructed as follows:

(1) In the bulk region ($0\leq i\leq j\leq N_{p}-3$): $U^{\left(1,0\right)}$
and $U^{\left(2,0\right)}$ as well as $U^{\left(0,1\right)}$ and
$U^{\left(0,2\right)}$ are computed  with  five points.
$U^{\left(1,1\right)}$ is computed with the nine points $\tilde{U}_{t}\left(i,j\right)$,
$\tilde{U}_{t}\left(\left(i\pm1\right),\left(j\pm1\right)\right)$,
$\tilde{U}_{t}\left(\left(i\pm1\right),\left(j\mp1\right)\right)$
$\tilde{U}_{t}\left(\left(i\pm2\right),\left(j\pm2\right)\right)$
and $\tilde{U}_{t}\left(\left(i\pm2\right),\left(j\mp2\right)\right)$.
The formulae are exact up to $\left(\Delta\psi\right)^{3}$.
Notice that for points on the two borders of $D$ defined either
by $\tilde{\psi}_{1}=0$ or $\tilde{\psi}_{1}=\tilde{\psi}_{2}$,
the derivatives of $\tilde U$ involve points outside $D$. By using (\ref{constraint}),
we can compute these values of $\tilde U$ from those that are inside $D$. This is one 
of the advantage of the choice of variables $(\psi_{1},\psi_{2})$ compared
to the choice $(\rho, \tau)$: The derivatives on the two borders 
$\tilde{\psi}_{1}=0$ and $\tilde{\psi}_{1}=\tilde{\psi}_{2}$ can be computed
in the same way as in the bulk.

(2) On the boundary of the domain $D$ corresponding to the large
field region, $j=N_{p}-2,N_{p}-1$, we compute the derivatives in the $\psi_1$ direction
$U^{\left(1,0\right)}(i,j)$
and $U^{\left(2,0\right)}(i,j)$ in the same way as in (1), that is, as in the bulk. The formulae
for $U^{\left(0,1\right)}(i,j)$ and $U^{\left(0,2\right)}(i,j)$
are constructed with the five quantities $\tilde{U}_{t}\left(i,j'\right)$
for $j'=N_{p}-5,\cdots,N_{p}-1$ and are exact at order $\left(\Delta\psi\right)^{2}$.
The formula for $U^{\left(1,1\right)}\left(i,N_{p}-1\right)$ for
$0\leq i\leq N_{p}-2$ involves the six values $\tilde{U}_{t}\left(i+1,j'\right)$,
$\tilde{U}_{t}\left(i-1,j'\right)$ for $j'=N_{p}-3,N_{p}-2,N_{p}-1$
and is exact at order $\left(\Delta\psi\right)$. Finally, for $U^{\left(1,1\right)}\left(N_{p}-1,N_{p}-1\right)$
we use twelve points in the region $N_{p}-4\leq i\leq j\leq N_{p}-1$
and the formula is exact at order $\left(\Delta\psi\right)^{2}$.

Notice that we have increased the precision
of the derivatives on the boundary of the domain $D$ corresponding
to the large field region in order to test the robustness of our results with respect
to the choice of discretization and to try to reduce numerical problems when $d$ is close
to 2. In all cases studied we did not find any significant changes. In particular,
the scheme is not more stable when the number of points chosen to compute the derivatives is increased.

Once the derivatives are discretized, the fixed point equation
$\partial_{t}\tilde{U}^{*}\left(\psi_{1},\psi_{2}\right)=0$
becomes a set of coupled algebraic equations for $g_{i,j}^{*}\equiv\tilde{U}\left(i,j\right)$.
We look for a solution to these equations by a Newton's-like method. One of the difficulty of this method
is the huge number of unknowns and the possibility for Newton's method
to get lost in the very complicated landscape of extrema of the set
of equations to be solved. The way out of this difficulty is to deform continuously
a solution of the problem.

Our strategy in this paper is to follow the fixed point potential
$\tilde{U}^{*}\left(\tilde{\psi}_{1},\tilde{\psi}_{2}\right)$ by changing
the dimension $d$ and the number of spin components $N$ gradually
starting from $d=3.9$ and $N=22$ where the
field-expansion method provides a good approximation of the fixed
point potential. We use as an initial condition of  Newton's method: 
\begin{equation}
\tilde{U}^{*,\,{\rm init}}\left(\tilde{\psi}_{1},\tilde{\psi}_{2}\right)=
\frac{\tilde{\lambda}^{*}}{2}\left(\tilde{\rho}-\tilde{\kappa}^{*}\right)^{2}+
\tilde{\mu}^{*}\tilde{\tau}
\label{eq:field expansion}
\end{equation}
and $\eta=0$.
The parameters  $\tilde{\lambda}^{*}$, $\tilde{\kappa}^{*}$ and $\tilde{\mu}^{*}$
are  determined by performing a field-expansion of the LPA equation on $\tilde{U}$ 
at order four in the fields and solving the fixed point equation 
for these parameters in $d=3.9$ and for $N=22$. As expected, we find four fixed points: the Gaussian and 
the $O(2N)$ fixed points as well as a once-unstable fixed point $C_{+}$ driving the phase transition
and $C_{-}$ that corresponds to a tricritical fixed point. Once an approximation of $C_{+}$ is found 
with the truncation of Eq. (\ref{eq:field expansion}), we use it as the initial condition of Newton's method
for the full potential equation (supplemented by $\eta$) and we easily find $\tilde{U}^*$.
Then, we move in the $(d,N)$ plane by little steps using as new initial condition what was found
for the previous value of $d$ and/or $N$ studied. The fixed potential potential deforms smoothly and 
the Newton's method always works properly this way.

\subsection{The line $N_c(d)$}

The line $N_c(d)$ separates in the $(d,N)$ plane the region where the phase transition 
is of second order and the region where it is of first order. When $N$ is lowered at fixed $d$, 
this line  corresponds to the locus of points  where $C_{+}$ disappears by collapsing with $C_{-}$.
There are two possibilities to determine $N_c(d)$. Either we decrease $N$ at fixed $d$ and look for
the value of $N$ where $C_{+}$ is no longer found and then repeat the same procedure by decreasing $d$.
Or we compute the smallest eigenvalue of the flow around the fixed point $C_{+}$ corresponding
to an irrelevant direction and look for the value of $N$ where it vanishes. This eigenvalue is 
a measure of the speed of the flow on the RG trajectory joining $C_{+}$ and $C_{-}$ and
this speed goes to 0 when the fixed points collapse. This second method is much more accurate
and less demanding than the first one and we therefore use it.

For each value of $(d,N)$ studied, we thus compute the eigenvalues of the stability
matrix $\Theta\left(\left\{ i,j\right\} ,\left\{ i',j'\right\} \right)$
defined as 
\begin{equation}
\Theta\left(\left\{ i,j\right\} ,\left\{ i',j'\right\} \right)
\equiv\frac{\partial\left(\partial_{t}g_{\left\{ i,j\right\} }\left(t\right)\right)}
{\partial g_{\left\{ i',j'\right\} }\left(t\right)}|_{g_{i,j}^{*}}
\end{equation}
 where we consider $\left\{ i,j\right\} $ and $\left\{ i',j'\right\} $
as (super-)indices. Since the RG time $t=\log k/\Lambda$ is negative,
a negative (positive) eigenvalue of the matrix $\Theta$ corresponds
to a relevant (irrelevant) eigendirection around the fixed point. We sort
the eigenvalues as $\sigma_{0}\left(=-d\right)<\sigma_{1}<\cdots<\sigma_{i-1}<\sigma_{i}<\cdots$.
Note that the above stability matrix around any fixed point solution
has a trivial relevant eigendirection corresponding to the constant shift $g_{i,j}=g_{i,j}^{*}+\mathrm{const}$
with the eigenvalue $\sigma_{0}=-d$, which can be easily seen from
Eq. (\ref{eq:flowU}). Hereafter, this trivial eigenvalue
is omitted when we discuss the stability of a fixed point. The critical
exponent $\nu$ is given by $\nu=-1/\sigma_{1}$ and the smallest positive 
eigenvalue we are interested in is $\sigma_{2}$.

\subsection{Numerical instabilities}
\begin{table}
\begin{equation}
\begin{array}{c|c}
N_p=61  &   -3, -1.45, 0.218, 0.827, 1.99, 2.79\\
        &      -0.464\pm 34.8i, 0.250\pm 30.9i, 1.07\pm 27.6i\\
N_p=81  &   -3 ,-1.45, 0.218, 0.827, 1.99, 2.79\\
        &       0.059 \pm 47.9i, 0.868 \pm 43.5i, 1.76 \pm 39.9 i\\
N_p=101 &   -3, -1.45, 0.218, 0.827, 1.99, 2.79,\\
        &       0.704\pm 61.05i,1.627\pm 56.3i\\
\end{array}
\end{equation}
\caption{Several of the most relevant eigenvalues around
the $C_{+}$ fixed point for $N=5$
and $d=3$.  The minimum of the potential corresponds to $\tilde{\psi}_{min}=3.96$ 
and we have chosen $\tilde{\psi}_{max}=9$. The physical eigenvalues are given on the first line 
for each value of $N_p$
and the others, that are spurious, on the second line. For $N_p=61$, the eigenvalues  $-0.464\pm 34.8i$
are relevant since their real part is negative. This eigenvalue disappears when increasing $N_p$.}
\label{eigenvalues} 
\end{table}
For each dimension $d$ and value of $N$ we have to make sure that our results are converged.
Once the choice of discretization of the derivatives has been made, there are two parameters that
can be tuned: the values of $\tilde{\psi}_{max}$ and of the mesh size
$\Delta\tilde{\psi}=\tilde{\psi}_{max}/\left(N_{p}-1\right)$. The potential $\tilde{U}^*$ shows
a minimum at $\tilde{\psi}_1=\tilde{\psi}_2=\tilde{\psi}_{min}$ and we have observed that $\tilde{\psi}_{max}$
should be at least 1.5 times larger than $\tilde{\psi}_{min}$ to get values of $N_c(d)$ converged with an accuracy
of less than $1\%$. We have also observed that the smaller  the 
dimension, the smaller  $\Delta\tilde{\psi}$ must be to get converged results. 
This last point has two origins. First, at small $d$ the fixed point potential is steep at large fields because
it behaves as $\left(\tilde{\psi}_{1}^{2}+\tilde{\psi}_{2}^{2}\right)^{\frac{d}{d-2+\eta}}$ and a small mesh size
is necessary to accurately describe the shape of $\tilde{U}^*$. Second, if $N_p$ is too small, 
we find that even far away from $d=2$,
say $d=3$,  several eigenvalues corresponding to relevant eigendirections appear in the spectrum and
spoil the degree of stability of the fixed point $C_+$. These eigenvalues
are clearly spurious because their values change considerably when either $\Delta\tilde{\psi}$
is decreased or $\tilde{\psi}_{max} $ is increased whereas the complementary set
of eigenvalues, the physical ones,  remain unchanged up to the sixth digit, see Table \ref{eigenvalues}.
We observe that 
as $\Delta\tilde{\psi}$ is decreased, these spurious eigenvalues systematically disappear (or, at least, get
a very large real part which makes them highly irrelevant). 
The conclusion of this study is that
for each $d$, a sufficiently large $N_p$ should be chosen so that the set of first most relevant eigenvalues
is converged as for their numbers and values. We find that in $d=3$, $N_p=101$ is sufficient to get fully
converged results while leading to numerically feasible calculations. We also find that as $d$ approaches 2,  
``large'' values
of $\tilde{\psi}_{max}$ favor the presence of spurious eigenvalues that can only be eliminated by
increasing $N_p$. It turns out that around $d=2.4$, very large values of $N_p$, such as $N_p=200$,
would be necessary to avoid spurious eigenvalues and that decreasing $d$ would impose to increase $N_p$
in a prohibitive way. We have been able to compute $N_c(d)$ down to $d=2.2$ by computing directly
the value of $N$ where no fixed point $C_+$ is found with Newton's method
but we have not been able to go below this dimension.

\section{Numerical results and conclusion}

\begin{figure}[t!]
	\centering 
    \includegraphics[width=0.65\textwidth]{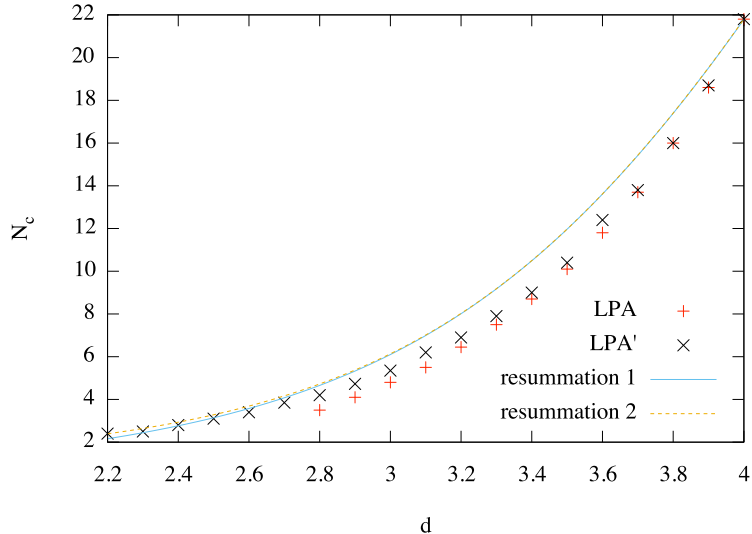}
    \caption{The curve $N_c(d)$. The crosses correspond to the calculation performed in this article either
    with the LPA or LPA'. The two continuous curves correspond to the five-loop results \cite{Calabrese five-loop} obtained within the
    $\epsilon$-expansion resummed either by assuming that $N_c(d=2)=2 $ in resummation 2 or assuming nothing about
    the value of $N_c(d=2)=2 $ in resummation 1.}
    \label{Nc}
\end{figure}

We have checked by varying all parameters ($\tilde\psi_{max}$ and $\Delta\psi$) that our results are 
 fully converged in $d=3$ from a numerical viewpoint both at the LPA and LPA' levels. They are also 
 converged down to $d\simeq 2.4$ and are less reliable in $d=2.2$ at the LPA' level although we are not able to give a quantitative
 estimate of the impact of our numerical errors on the value of $N_c(d)$ in this dimension. We show our 
 determination of $N_c(d)$  in Fig. \ref{Nc} together with the results obtained from the $epsilon$-expansion
 at five loops. 
 
For $d=3$, our results confirm the previous results obtained either by the NPRG \cite{Zumbach1, Zumbach2, NPRG field expansion, Delamotte Review, NPRG semi-expansion} 
or the $\epsilon$-expansion approaches \cite{Jones, Bailin, Calabrese five-loop}. 
The comparison between the LPA and LPA' results
strongly suggests that neglecting the effect of the derivative terms on the determination of $N_c(d)$ plays
a minor role in $d=3$. Moreover, $N_c(d=3)$ increases
between the LPA and LPA' and becomes closer to the results obtained with the $\epsilon$-expansion, which is
expected. It seems  therefore very difficult to imagine that $N_c(d=3)$ could be smaller than 3.

Let us also emphasize that the only Monte Carlo simulation that still finds a second order transition for a value of
$N$ below our value of $N_c(d=3)$, that is, for $N\le4$, has been performed for $N=2$ by Calabrese et al. \cite{Calabrese 3} on a discretization
of the Ginzburg-Landau model Eqs.~(\ref{eq:effective hamiltonian}), (\ref{pot}). They found that depending on the values of $\lambda$ and $\mu$,
the transition is of first or second order: At fixed $\lambda$ and small $\mu$, the transition is of second order whereas it is of first order
at large $\mu$. Since nonuniversal quantities, such as  phase diagrams \cite{machado,Canet04}, can be accurately 
computed from the integration of the NPRG flow equations, 
it is possible to estimate the magnitude of the correlation length $\xi_c$ at the  transition within the LPA' 
by initializing the flow with the data corresponding to the simulations. By varying these data as well as the cut-off function
$R_k(q)$, it is found that $\xi_c$ is always  finite  (since there is no fixed point) but very large, typically larger
than 2000 lattice spacings \cite{Debelhoir}. From a numerical point of view, there is no doubt that  such 
a large correlation length makes impossible to decide in favor of a second or a (very weak) first order phase transition since in both cases the physics 
will look the same at the scale of the lattice size which was at most 120 lattice spacings in the numerical simulations. 
We conclude that this Monte Carlo result does not
contradict our conclusion that $N_c(d=3)\simeq 5$.

This result shows unambiguously that if our result is wrong, the origin of the problem can only be
found by including the renormalization of the functions in front of the derivative terms. However,
considering that the 
anomalous dimension is small for these systems when they undergo a second order phase transition, that is,
for $N>N_c$, this
hypothesis seems very doubtful. We therefore suggest that it is useless to study the order two of the derivative 
expansion in these models that, most probably, would bring only minor modifications as compared to the present
study. We also suggest that only the Blaizot-Mendez-Wschebor approach \cite{BMW1,BMW2,BMW3}, where the full momentum dependence of the 
two-point functions is retained as well as the full field-dependence of the potential $\tilde{U}$ could lead
to a very accurate determination of $N_c(3)$.

As for the approach to $d=2$, we find a remarkable agreement between our results and what was found within
the $\epsilon$-expansion.
Two resummations of the $\epsilon$-expansion were performed by the authors of \cite{Calabrese five-loop},
either by assuming that $N_c(d=2)=2 $ or by letting free
the value of $N_c(d=2) $. This agreement is not very surprising because we expect  the LPA' to be
accurate around $d=2$ for $N>3$ (it is one-loop exact in the nonfrustrated case). Notice that our results are not precise enough to determine unambiguously
the value of $N_c(d=2) $
although it seems clear that it cannot be very different from 2.
It is therefore very unlikely that $N_c(d=2)>3$ and our results show that the $C_+$ fixed point must exist
for all dimensions larger than two in the Heisenberg case. Since the NPRG flow reproduces the low-temperature
expansion of the nonlinear sigma model around $d=2$, we conclude that the critical behavior of frustrated
systems in $d=2+\epsilon$ is driven for $N=3$ by the fixed point $C_+$ corresponding to a critical temperature
of order $\epsilon$ in agreement with Mermin-Wagner theorem. Since we find no other once-unstable
fixed point, we conclude that our study rules out the possibility of having a finite temperature
fixed point in $d=2$ for $N=3$ contrary to what was found at five loops in a fixed
dimension RG calculation \cite{Calabrese focus fixed points 2}.

To conclude, we have presented a rather simple method to compute the fixed point properties 
of matrix models describing frustrated systems without having recourse to a field expansion
of the free energy $\Gamma$ (but keeping a derivative expansion of $\Gamma$). This is especially
important in low dimensions where the field expansion is known to fail. In dimension $d=3$, our results
fully confirm what was previously found within less accurate NPRG calculations that involved 
 field truncations on top of the derivative expansion \cite{NPRG field expansion, NPRG semi-expansion, Delamotte Review}. In dimension $d=2$, more stable numerical
 schemes are still needed to study the physics of topological excitations in frustrated systems
 (that are of different natures than in nonfrustrated systems) and we believe that the present work
 is the first step in this direction.

\section{Acknowledgment}

This work was supported in part by a Grant-in-Aid for Young
 Scientists
(B) (15K17737), Grants-in-Aid for Japan Society for Promotion of Science (JSPS) Fellows (Grants Nos. 241799 and 263111), the JSPS Core-to-Core Program "Non-equilibrium dynamics of soft matter and information".

\appendix

\section{The nonperturbartive renormalization group flow equations and the anomalous dimension}

Throughout this paper we employ the following $R_{k}\left(\mathbf{q}^{2}\right)$,
which is useful for analytical treatments\cite{Litim}: 
\begin{equation}
R_{k}\left(\mathbf{q}^{2}\right)=Z_{k}\left(k^{2}-\mathbf{q}^{2}\right)\Theta\left(k^{2}-\mathbf{q}^{2}\right),
\end{equation}
where $Z_{k}$ is defined as 
\begin{equation}
Z_{k}=\left(\frac{\partial}{\partial p^{2}}\left(\frac{\delta^{2}\Gamma_{k}}{\delta\phi_{1}^{3}\left(\mathbf{p}\right)\delta\phi_{1}^{3}\left(-\mathbf{p}\right)}/\left(2\pi\right)^{d}\delta\left(\mathbf{0}\right)\right)\right)_{\mathbf{p}=0,min},
\end{equation}
where the field values are set to the minimum of $U_{k}$ given by Eq. (\ref{eq:min}). Here the Fourier transform $\phi_{1}^{3}\left(\mathbf{p}\right)$
is defined as $\phi_{1}^{3}\left(\mathbf{p}\right)=\int d^{d}\mathbf{x}\phi_{1}^{3}\left(\mathbf{x}\right)\exp\left(-i\mathbf{x}\cdot\mathbf{q}\right).$

Then, the running anomalous dimension $\eta_{k}=-k\partial_{k}Z_{k}$
is given, at the level of LPA', by

\begin{eqnarray}
\eta_{k} & = & 64\frac{\tilde{\kappa}v_{d}}{d}\left(\frac{1}{1+2\tilde{U_{k}}^{\left(1,0\right)'}}\right)^{2}\nonumber \\
 &  & \times\left(2\left(\frac{\tilde{U}_{k}^{\left(2,0\right)'}}{1+2\tilde{U_{k}}^{\left(1,0\right)'}+4\tilde{\kappa}\tilde{U}^{\left(2,0\right)'}}\right)^{2}+\left(\frac{\tilde{U}_{k}^{\left(0,1\right)'}}{1+2\tilde{\kappa}\tilde{U}_{k}^{\left(0,1\right)'}+2\tilde{U_{k}}^{\left(1,0\right)'}}\right)^{2}\right),\nonumber \\
\label{eq:eta}
\end{eqnarray}
where we set $\tilde{\rho}=\tilde{\kappa}$ and $\tilde{\tau}=0$.
The derivatives $\tilde{U}_{k}^{\left(i,j\right)'}$ with respect
to the invariants $\tilde{\rho}$ and $\tilde{\tau}$, and $v_{d}$
are defined as 
\[
\tilde{U}_{k}^{\left(i,j\right)'}\equiv\frac{\partial^{i+j}\tilde{U}_{k}}{\partial\tilde{\rho}^{i}\partial\tilde{\tau}^{j}},v_{d}=\frac{1}{2^{d+1}\pi^{d/2}\Gamma\left(\frac{d}{2}\right)}.
\]

The scaled nonperturbartive renormalization group flow equation for the porential $\tilde{U}_{k}$ is
given by
\begin{eqnarray}
\partial_{t}\tilde{U}_{k} & = & -d\tilde{U}_{k}+\frac{1}{2}(-2+d+\eta_{k})\left(\tilde{\psi}_{1}\tilde{U}_{k}^{\left(1,0\right)}+\tilde{\psi}_{2}\tilde{U}_{k}^{\left(0,1\right)}\right)\nonumber \\
 &  & +\frac{4(2+d-\eta_{k})}{d(2+d)}v_{d}\nonumber \\
 &  & \times\left(\frac{\tilde{\psi}_{1}-\tilde{\psi}_{2}}{\tilde{\psi}_{1}-\tilde{\psi}_{2}-\tilde{U}_{k}^{\left(0,1\right)}+\tilde{U}_{k}^{\left(1,0\right)}}+\frac{\tilde{\psi}_{1}+\tilde{\psi}_{2}}{\tilde{\psi}_{1}+\tilde{\psi}_{2}+\tilde{U_{k}}^{\left(0,1\right)}+\tilde{U}_{k}^{\left(1,0\right)}}\right.\nonumber \\
 &  & +(N-2)\left(\frac{\tilde{\psi}_{2}}{\tilde{\psi}_{2}+\tilde{U}_{k}^{\left(0,1\right)}}+\frac{\tilde{\psi}_{1}}{\tilde{\psi}_{1}+\tilde{U}_{k}^{\left(1,0\right)}}\right)\nonumber \\
 &  & \left.+{\normalcolor }\frac{2+\tilde{U_{k}}^{\left(0,2\right)}+\tilde{U}_{k}^{\left(2,0\right)}}{1-\left(\tilde{U}_{k}^{\left(1,1\right)}\right)^{2}+\tilde{U}_{k}^{\left(2,0\right)}+\tilde{U}_{k}^{\left(0,2\right)}\left(1+\tilde{U}_{k}^{\left(2,0\right)}\right)}\right).\label{eq:flow eq}
\end{eqnarray}
Here, to simplify the notation, we have defined another kind of derivatives $\tilde{U}_{k}^{\left(i,j\right)}$ with respect
to $\tilde{\psi}_{1}$ and $\tilde{\psi}_{2}$ as
\begin{equation}
\tilde{U}_{k}^{\left(i,j\right)}\equiv\frac{\partial^{i+j}\tilde{U}_{k}}{\partial\tilde{\psi}_{1}^{i}\partial\tilde{\psi}_{2}^{j}}.
\label{eq:flowU}
\end{equation}

In our calculations, we use the rescaled potential ${v_{d}^{-1}\tilde{U}_{k}}$ and fields $\left(v_{d}\right)^{-1/2}\tilde{\psi}_{i}$
for $i=1,2$ 
in such a way that $v_{d}$ disappears in Eqs (\ref{eq:eta}) and
(\ref{eq:flow eq}).

\end{document}